\documentclass[journal]{IEEEtran}

\ifCLASSINFOpdf
  \usepackage[pdftex]{graphicx}
  \DeclareGraphicsExtensions{.pdf,.jpeg,.png}
\else
  \usepackage[dvips]{graphicx}
\fi
\usepackage{subfigure}
\usepackage{epstopdf}
\usepackage[cmex10]{amsmath}
\usepackage{amssymb}
\usepackage{wasysym}
\usepackage{algorithm}
\usepackage{algorithmic}
\usepackage{array}
\usepackage{cite}
\usepackage{color}
\usepackage{url}
\usepackage{bm}
\usepackage{enumerate}
\usepackage{epsfig,latexsym}
\usepackage{flushend}
\usepackage{cite}
\usepackage{hyperref}
\usepackage{diagbox} 
\usepackage{booktabs}

\begin{document}
%
\title{Hierarchical Codebook based Multiuser Beam Training for Millimeter Wave Massive MIMO}

\author{Chenhao~Qi,~\IEEEmembership{Senior~Member,~IEEE}, Kangjian~Chen,~\IEEEmembership{Student~Member,~IEEE}, \\ Octavia A. Dobre,~\IEEEmembership{Fellow,~IEEE},  and Geoffrey Ye Li,~\IEEEmembership{Fellow,~IEEE}
\thanks{This paper was presented in part at 2019 IEEE Global Communications Conference (GLOBECOM), Waikoloa Village, HI, USA, December 2019~\cite{chen2019simultaneous}. This work was supported in part by National Natural Science Foundation of China under Grant 61871119, by Natural Science Foundation of Jiangsu Province under Grant BK20161428, and by the Natural Sciences and Engineering Research Council of Canada (NSERC) through its Discovery Program. (\textit{Corresponding author: Chenhao~Qi})}
\thanks{Chenhao~Qi and Kangjian~Chen are with the School of Information Science and Engineering, Southeast University, Nanjing 210096, China (Email: qch@seu.edu.cn).}
\thanks{Octavia A. Dobre is with the Faculty of Engineering and Applied Science, Memorial University, Canada (Email: odobre@mun.ca).}
\thanks{Geoffrey Ye Li is with the School of Electrical and Computer Engineering, Georgia Institute of Technology, Atlanta, GA, USA (Email: liye@ece.gatech.edu).}
}

\markboth{Accepted By IEEE Transactions on Wireless Communications}
{Shell \MakeLowercase{\textit{et al.}}: Bare Demo of IEEEtran.cls for Journals}

\maketitle

\begin{abstract}
In this paper, multiuser beam training based on hierarchical codebook for millimeter wave massive multi-input multi-output is investigated, where the base station (BS) simultaneously performs beam training with multiple user equipments (UEs). For the UEs, an alternative minimization method with a closed-form expression (AMCF) is proposed to design the hierarchical codebook under the constant modulus constraint. To speed up the convergence of the AMCF, an initialization method based on Zadoff-Chu sequence is proposed. For the BS,  a simultaneous multiuser beam training
scheme based on an adaptively designed hierarchical codebook is proposed, where the codewords in the current layer of the codebook are designed according to the beam training results of the previous layer. The codewords at the BS are designed with multiple mainlobes, each covering a spatial region for one or more UEs. Simulation results verify the effectiveness of the proposed hierarchical codebook design schemes and show that the proposed multiuser beam training scheme can approach the performance of the beam sweeping but with significantly reduced beam training overhead.
\end{abstract}
\begin{IEEEkeywords}
Beam training, hierarchical codebook, massive multi-input multi-output (MIMO), millimeter wave (mmWave) communications.
\end{IEEEkeywords}

\section{Introduction}
Millimeter wave (mmWave) massive multi-input multi-output (MIMO) has been considered as a promising technology for future wireless communications due to its rich spectral resource~\cite{heath2016overview,ZhaoLou,LiYemmWave,xiao2018noma,Wei2019multi,BLWandYeli}. However, the transmission of mmWave signal experiences large path loss because of its high frequency. To compensate it, antenna arrays with a hybrid precoding architecture have been introduced, where a small number of radio frequency (RF) chains are connected to a large number of antennas via phase shifters~\cite{Ma2020Sparse}. Compared to the fully digital architecture, the hybrid precoding architecture can substantially reduce the hardware complexity and save the energy consumption.

To acquire the channel state information (CSI) in mmWave massive MIMO systems with the hybrid precoding structure, codebook-based beam training methods have been widely adopted~\cite{CommonCodebookDesign,TSPMKok,AlMultiUser,TWCSXY}. To reduce the training overhead, hierarchical codebook-based beam training methods have been proposed. For hierarchical beam training, a predefined hierarchical codebook including several layers of codebooks is typically employed, where the spatial region covered by the codeword at an upper layer of the codebook is split into several smaller spatial regions covered by codewords at a lower layer~\cite{Xiao2016Hierarchical}. Earlier work on hierarchical beam training focuses on peer-to-peer mmWave massive MIMO systems. In \cite{Sparse2014}, a hierarchical codebook is utilized to acquire the CSI in mmWave massive MIMO systems, where channel estimation is formulated as a sparse reconstruction problem. For multiuser scenarios, a straightforward extension of the above work is time-division multiple access (TDMA) hierarchical beam training, where the base station (BS) sequentially performs the hierarchical beam training user by user and each occupies a different part of time. However, the total training overhead grows linearly with the number of users. To reduce the overhead of beam training, a simultaneous hierarchical beam training for a multiuser mmWave massive MIMO system is proposed for partially connected structure, where each RF chain is solely connected to an antenna subarray at the BS and the beam training is independently performed by each subarray~\cite{SC-2018}.

\begin{figure*}[htbp]
  \centering
  \includegraphics[width=0.8\textwidth]{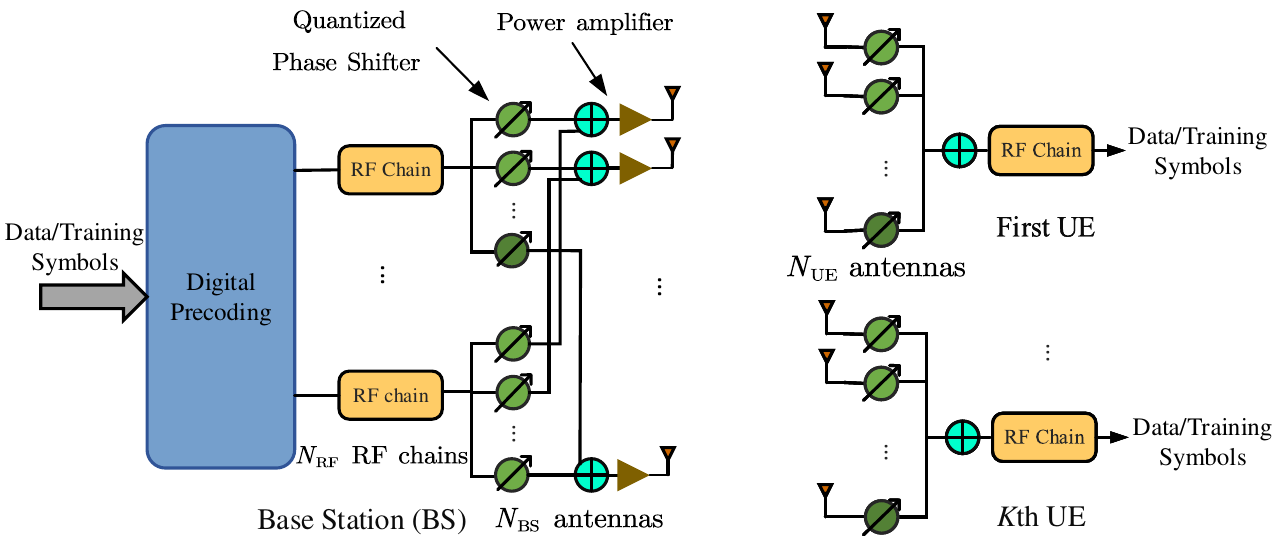}
  \caption{Illustration of a multiuser mmWave massive MIMO system with a BS and $K$ UEs.}
  \label{fig:system model}
\end{figure*}

So far, several methods have been proposed to design the hierarchical codebook~\cite{ChenhaoQiOverview}. Many methods exploit the degree of multiple RF chains to construct the codebook. If there is only one RF chain which is the general setup for most UEs, the following two methods can be employed to design the hierarchical codebook. In \cite{Xiao2016Hierarchical}, a joint sub-array and de-activation (JOINT) method is proposed, where the weighted summation of several sub-arrays is used to form wide beams during codebook design. However, half of the antennas may be powered off for the JOINT method, which will weaken the signal strength and thus reduce the coverage. To address this issue, an enhanced JOINT (EJOINT) method is developed, where the codewords are formed without antenna de-activation~\cite{ZhenyuXiao2018}.

In this paper, we propose a simultaneous multiuser hierarchical beam training scheme based on our designed adaptive hierarchical codebook for multiuser mmWave massive MIMO systems. The main contributions of this paper are as follows.

\begin{enumerate}
\item For the UEs served by the BS, we propose an alternative minimization method with a closed-form expression (AMCF) to design the hierarchical codebook under the constant modulus constraint. To speed up the convergence of the AMCF, one initialization method is proposed, namely AMCF with Zadoff-Chu sequence initialization (AMCF-ZCI).
\item For the BS, rather than sequentially performing the beam training with different users using the same hierarchical codebook, we adaptively design the hierarchical codebook, where the codewords in the current layer of the hierarchical codebook are determined by the beam training results of the previous layer. In particular, we design multi-mainlobe codewords for the BS to simultaneously perform the beam training with all UEs, where each mainlobe covers a spatial region for one or more UEs. Excluding the bottom layer of the hierarchical codebook, there are only two codewords at each layer in the designed adaptive hierarchical codebook, which only requires two times of simultaneous beam training for all the UEs no matter how many UEs the BS serves. Compared with the existing beam training schemes, the proposed simultaneous multiuser hierarchical beam training scheme can substantially reduce the training overhead.
\end{enumerate}

The rest of this paper is organized as follows. The problem of multiuser beam training is formulated in Section~\ref{Sec.Problem Formulation}. The codebook design for the UEs is investigated in Section~\ref{Sec.UE}. The codebook design for the BS and the simultaneous beam training is discussed in Section~\ref{Sec.BS}. Simulation results are provided in Section~\ref{Sec.Simulation}. Finally, Section~\ref{Sec.Conclusion} concludes this paper.

The notations are defined as follows. Symbols for matrices (upper case) and vectors (lower case) are in boldface. $[ \boldsymbol{a} ] _n$, $[ \boldsymbol{A} ] _{:,n}$ and $[ \boldsymbol{A} ] _{m,n}$ denote the $n$th entry of a vector $\boldsymbol{a}$, the $n$th column of a matrix $\boldsymbol{A}$, and the entry on the $m$th row and $n$th column of $\boldsymbol{A}$. According to the convention, $\boldsymbol{I}$, $(\cdot)^T$, $(\cdot)^H $, $\|\cdot \|_F$, $\|\cdot \|_2$, $\mathbb{C}$, $\mathbb{E}\{ \cdot \}$, $\rm{diag}\{ \cdot \}$, $\circ$ and $\mathcal{C}\mathcal{N}$ denote the identity matrix, transpose, conjugate transpose (Hermitian), Frobenius norm, $\ell_2$-norm, set of complex number, operation of expectation, diagonal matrix, entry-wise product and complex Gaussian distribution, respectively. $\angle(\cdot)$ denotes the phase of a complex value or a vector.

\section{System Model}\label{Sec.Problem Formulation}
As shown in Fig.~\ref{fig:system model}, we consider a multiuser mmWave massive MIMO system with a BS and $K$ UEs. The number of antennas at the BS and each UE is $N_{\rm BS}$ and $N_{\rm UE}(N_{\rm UE}\leq N_{\rm BS})$, respectively. The number of RF chains at the BS and each UE is $N_{\rm RF}(N_{\rm RF}\ll N_{\rm BS} )$ and one, respectively. To simplify the analysis, both $N_{\rm BS}$ and $N_{\rm BS}$ are set as an integer power of two. The BS employs hybrid precoding, including digital precoding and analog precoding while each UE employs analog combining. At the BS, each RF chain is connected to $N_{\rm BS}$ antennas via $N_{\rm BS}$ quantized phase shifters. At each UE, one RF chain is connected to $N_{\rm UE}$ antennas via $N_{\rm UE}$ quantized phase shifters. The antennas at both the BS and the UEs are placed into uniform linear arrays (ULAs) with half wavelength spacing. Generally, each RF chain at the BS can support an independent data stream for a UE. Therefore the number of UEs simultaneously served by the BS is usually smaller than the number of RF chains, i.e., $K\le N_{\rm RF}$.

During the downlink signal transmission from the BS to the UEs, the received signal by the $k$th UE for $k=1,2,\ldots,K$ can be expressed as
\begin{equation}\label{Received Signal}
  y_k=\boldsymbol{w}_k^H\boldsymbol{H}_k\boldsymbol{F}_{\rm RF}\boldsymbol{F}_{\rm BB} \boldsymbol{P}\boldsymbol{s}+\boldsymbol{w}_k^H\boldsymbol{n}_k,
\end{equation}
where $y_k$, $\boldsymbol{w}_k\in \mathbb{C}^{N_{\rm UE}}$, $\boldsymbol{H}_k\in \mathbb{C}^{N_{\rm UE}\times N_{\rm BS}}$, $\boldsymbol{F}_{\rm RF}\in \mathbb{C}^{N_{\rm BS}\times N_{\rm RF}}$, $\boldsymbol{F}_{\rm BB}\in \mathbb{C}^{N_{\rm RF}\times K}$, $\boldsymbol{P}\triangleq {\rm{diag}}\{\sqrt{P_1}, \sqrt{P_2}, \ldots, \sqrt{P_K}\}$, $\boldsymbol{s}\in \mathbb{C}^{K}$ and $\boldsymbol{n}_k\in \mathbb{C}^{N_{\rm UE}}$ denote the received signal, the analog combiner of the $k$th UE, the channel matrix between the BS and the $k$th UE, the analog precoder of the BS, the digital precoder of the BS, the diagonal power allocation matrix, the transmitted signal vector, and the additive white Gaussian noise (AWGN) vector obeying $\boldsymbol{n}_k\sim\mathcal{CN}(0,\sigma^2\boldsymbol{I}_{N_{\rm UE}})$, respectively. Note that the hybrid precoder, including the analog precoder and digital precoder, has no power gain, i.e.,  $\|\boldsymbol{F}_{\rm RF}\boldsymbol{F}_{\rm BB}\|_F^2=K$. The power allocation subjects to the constraint that $\sum_{k=1}^{K} P_k=P_{\rm Total}$, where $P_k$ denotes the power allocated to the $k$th UE for $k=1,2,\dots, K$. Moreover, the transmitted signal vector $\boldsymbol{s}$  subjects to the unit power constraint that $\mathbb{E}\{\boldsymbol{s}\boldsymbol{s}^H\}=\boldsymbol{I}_{K}$.

According to the widely used Saleh-Valenzuela channel model~\cite{heath2016overview,TSPMWY}, the mmWave MIMO channel matrix, $\boldsymbol{H}_k\in \mathbb{C}^{N_{\rm UE}\times N_{\rm BS}}$, between the BS and the $k$th UE can be expressed as
\begin{equation}\label{channel model}
\boldsymbol{H}_k=\sqrt{\frac{N_{\rm BS} N_{\rm UE}}{L_k}}
\sum_{l=1}^{L_k}\lambda_l\boldsymbol{\alpha}\big(N_{\rm UE},\theta_{\rm UE}^{l}\big)\boldsymbol{\alpha}^H\big(N_{\rm BS},\theta_{\rm BS}^{l}\big)
\end{equation}
where $L_k$, $\lambda_l$, $\theta_{\rm UE}^{l}$, and $\theta_{\rm BS}^{l}$ denote the number of multi-path, the channel gain, the channel angle-of-arrival (AoA), and the channel angle-of-departure (AoD) of the $l$th path, respectively. In fact, $\theta_{\rm UE}^{l}=\cos\big( \omega _{\rm UE}^{l} \big)$ and $\theta_{\rm BS}^{l}=\cos\big( \omega _{\rm BS}^{l} \big)$, where $\omega _{\rm UE}^{l}$ and $\omega _{\rm BS}^{l}$ denote the physical AoA and AoD of the $l$th path, respectively. Since $\omega _{\rm UE}^{l}\in \left[ 0,2\pi \right)$ and $ \omega _{\rm BS}^{l}\in \left[ 0,2\pi \right)$, we have $\theta_{\rm UE}^{l}\in [-1,1]$ and $\theta_{\rm BS}^{l} \in [-1,1]$. The channel steering vector in \eqref{channel model} is defined as
\begin{equation}\label{steering vector}
\boldsymbol{\alpha}\left( N,\theta \right) =\frac{1}{\sqrt{N}}\left[ 1,e^{j\pi \theta},\cdots ,e^{j\left( N-1 \right) \pi \theta} \right] ^T
\end{equation}
where $N$ is the number of antennas and $\theta$ is the channel AoA or AoD.

Our objective is to maximize the averaged sum-rate of $K$ UEs by adjusting $\boldsymbol{F}_{\rm RF}$, $\boldsymbol{F}_{\rm BB}$, $\{\boldsymbol{w}_k\}_{k=1}^K$ and $\{P_k\}_{k=1}^K$. It can be expressed as the following optimization problem,
\begin{subequations}\label{optimization problem}\normalsize
    \begin{align}
    \underset{\boldsymbol{F}_{\rm RF},\boldsymbol{F}_{\rm BB},\atop{\{\boldsymbol{w}_k,P_k\}_{k=1}^K}}{\max} & \sum_{k=1}^{K}~\frac{1}{K}R_k \label{Objective1}\\
\mathrm{s.t.}~~~~~&\big|\left[ \boldsymbol{F}_{\rm RF} \right] _{n,t}\big|=\frac{1}{\sqrt{N_{\rm BS}}}, \big|[\boldsymbol{w}_k]_m\big|=\frac{1}{\sqrt{N_{\rm UE}}}, \label{envelop constrain1}\\
& \big\| \boldsymbol{F}_{\rm RF} [\boldsymbol{F}_{\rm BB}]_{:,k}\big\|_2=1, \label{power constraint}\\
& P_1+P_2+\cdots+P_K=P_{\rm Total}, \label{power allocation}\\
    &n=1,2,\ldots,N_{\rm BS},~m=1,2,\ldots, N_{\rm UE},\\ \nonumber
    &t=1,2,\cdots,N_{\rm RF},~k=1,2,\cdots K. \nonumber
    \end{align}
\end{subequations}
where
\begin{equation}\label{sum rate}
R_k=\log_2\bigg(1+\frac{{P_k}\big|\boldsymbol{w}_k^H\boldsymbol{H}_k\boldsymbol{F}_{\rm RF}[\boldsymbol{F}_{\rm BB}]_{:,k}\big|^2}{\sum_{i\neq k} {P_i} \big|\boldsymbol{w}_k^H\boldsymbol{H}_k\boldsymbol{F}_{\rm RF}[\boldsymbol{F}_{\rm BB}]_{:,i}\big|^2+\sigma^2}\bigg)
\end{equation}
is the achievable rate of the $k$th UE for $k=1,2,\dots, K$. In the above, \eqref{envelop constrain1} indicates that the entries in $\boldsymbol{F}_{\rm RF}$ and $\boldsymbol{w}_k$ satisfy the constant envelop constraint of phase shifters, \eqref{power constraint} indicates that the hybrid precoder does not provide power gain, and \eqref{power allocation} indicates the power allocation for different users.

In~\eqref{sum rate}, $\boldsymbol{H}_k$ is required to calculate $R_k$. But the codebook-based beam training can avoid directly using $\boldsymbol{H}_k$, since the estimation of $\boldsymbol{H}_k$ incurs a large overhead. We denote the codebooks at the BS and each UE as $\boldsymbol{\mathcal{F}}=\{\boldsymbol{f}_{\rm c}^1,\boldsymbol{f}_{\rm c}^2,\cdots,\boldsymbol{f}_{\rm c}^{N_{\rm BS}}\}$ and  $\boldsymbol{\mathcal{W}}=\{\boldsymbol{w}_{\rm c}^1,\boldsymbol{w}_{\rm c}^2,\cdots,\boldsymbol{w}_{\rm c}^{N_{\rm UE}}\}$, respectively, where
\begin{align}\label{Codebook}
  &\boldsymbol{f}_{\rm c}^n=\boldsymbol{\alpha}(N_{\rm BS},-1+(2n-1)/{N_{\rm BS}}),\nonumber\\
  &\boldsymbol{w}_{\rm c}^m=\boldsymbol{\alpha}(N_{\rm UE},-1+(2m-1)/{N_{\rm UE}}).
\end{align}

The objective of beam training is to select $K$ codewords from $\boldsymbol{\mathcal{F}}$ for the BS and $K$ codewords from $\boldsymbol{\mathcal{W}}$ for $K$ UEs to maximize the averaged sum-rate of the $K$ UEs. Then \eqref{optimization problem} can be rewritten as
\begin{subequations}\label{optmization problem2}\normalsize
    \begin{align}
    \underset{\boldsymbol{F}_{\rm RF},\boldsymbol{F}_{\rm BB},\atop{\{\boldsymbol{w}_k,P_k\}_{k=1}^K}}{\max} & \sum_{k=1}^{K}~\frac{1}{K}R_k \label{Objective2}\\
\mathrm{s.t.}~~~~~&\boldsymbol{f}_k\in\boldsymbol{\mathcal{F}},\boldsymbol{w}_k\in \boldsymbol{\mathcal{W}},~\big\| \boldsymbol{F}_{\rm RF} [\boldsymbol{F}_{\rm BB}]_{:,k}\big\|_2=1, \label{envelop constrain2}\\
& P_1+P_2+\cdots+P_K=P_{\rm Total}, \label{power allocation2}\\
    &k=1,2,\cdots, K \nonumber
    \end{align}
\end{subequations}
where $\boldsymbol{f}_k \triangleq [\boldsymbol{F}_{\rm RF}]_{:,k}$. In \eqref{optmization problem2}, the coupling among $\boldsymbol{F}_{\rm RF}$, $\boldsymbol{F}_{\rm BB}$ and $\{P_k\}_{k=1}^K$ makes it difficult to solve the problem directly. A typical method in the existing work is to design $\boldsymbol{w}_k$ and $\boldsymbol{F}_{\rm RF}$ first without considering $\boldsymbol{F}_{\rm BB}$ and $\{P_k\}_{k=1}^K$, and then design $\boldsymbol{F}_{\rm BB}$ under the power constraint of (4c) to eliminate multiuser interference~\cite{TWCSXY}. Once the beam training is finished and the CSI is obtained, we design $\{P_k\}_{k=1}^K$ for data transmission, e.g., using the water-filling method. During the beam training, we do not consider the power allocation and use the equal power for different users.

Since designing $\boldsymbol{F}_{\rm RF}$ is essentially to find $\boldsymbol{f}_k$, our objective turns to find a pair of $\boldsymbol{f}_k$ and $\boldsymbol{w}_k$ best fit for $\boldsymbol{H}_k$, which can be expressed as
\begin{align}\label{analog optimization}
  \max_{ \boldsymbol{f}_k, \boldsymbol{w}_k} &|\boldsymbol{w}_k^H\boldsymbol{H}_k\boldsymbol{f}_k|\\
  {\rm s.t.}~~&\boldsymbol{f}_k \in \boldsymbol{\mathcal{F}},~\boldsymbol{w}_k \in \boldsymbol{\mathcal{W}}.\nonumber\nonumber
\end{align}
A straightforward method to solve \eqref{analog optimization} is the exhaustive beam training, which is also called beam sweeping. It tests all possible pairs of $\boldsymbol{f}_k$ and $\boldsymbol{w}_k$ to find the best one. However, such a method takes a long time and therefore with a large overhead. If we denote the period of each test of a pair of $\boldsymbol{f}_k$ and $\boldsymbol{w}_k$ as a time slot, the exhaustive beam training needs totally $N_{\rm BS}N_{\rm UE}$ time slots. Note that the number of total time slots is independent of $K$ since the UEs can simultaneously test the power of their received signal and eventually feed back the indices of the best codewords to the BS.


To reduce the overhead of exhaustive beam training, hierarchical beam training that is based on hierarchical codebooks, is widely adopted~\cite{Xiao2016Hierarchical}. The hierarchical codebook typically consists of a small number of low-resolution codewords covering a wide angle at the upper layer of the codebook and a large number of high-resolution codewords offering high directional gain at the lower layer of the codebook. The hierarchical beam training usually first tests the mmWave channel with some low-resolution codewords at the upper layer and then narrows down the beam width layer by layer until a codeword pair at bottom layer is obtained. We denote the hierarchical codebooks employed at the BS and UEs as $\boldsymbol{\mathcal{V}}_{\rm BS}$ and $\boldsymbol{\mathcal{V}}_{\rm UE}$, respectively. The $m$th codeword at the $s$th layer of $\boldsymbol{\mathcal{V}}_{\rm UE}$ for $s=1,\cdots,T$ and $m=1,2,\cdots2^s$ is denoted as $\boldsymbol{\mathcal{V}}_{\rm UE}(s,m)$ where
\begin{equation}
  T=\log_2{N_{\rm UE}}
\end{equation}
is the number of layers of $\boldsymbol{\mathcal{V}}_{\rm UE}$. Note that the codewords at the bottom layer of  $\boldsymbol{\mathcal{V}}_{\rm UE}$ are exactly the same as the codewords in $\boldsymbol{{\mathcal{W}}}$, which implies that the motivation of hierarchical beam training is to adopt the merits of binary tree to  improve the efficiency of beam training. Compared with the exhaustive beam training, the hierarchical beam training sequentially performed user by user in the TDMA fashion can reduce the overhead from $N_{\rm BS}N_{\rm UE}$ to $2K(\log_2N_{\rm BS}+\log_2 N_{\rm UE})$.  For example, if $K=4$, $N_{\rm BS}=64$ and $N_{\rm UE}=16$, the hierarchical beam training can reduce the training overhead by $92.2\%$ compared to the exhaustive beam training.

\section{Hierarchical Codebook Design for UEs} \label{Sec.UE}
In this section, we design the hierarchical codebook for the UEs, while the hierarchical codebook designed for the BS will be addressed in the next section together with the beam training.

\subsection{Codebook Design Based on Alternative Minimization Method}

Since each UE normally has only one RF chain, the $N_{\rm UE}$ antennas at each UE are connected to the RF chain via phase shifters without digital precoding. Therefore, each entry of the codeword in the hierarchical codebook must be designed under the constant modulus constraint. In the following we propose an alternative minimization method with a closed-form expression (AMCF).

Denote the absolute beam gain of a codeword $\boldsymbol{v}\in \mathbb{C}^{N_{\rm UE}}$ with beam coverage $\mathcal{I}_v=[\Omega_0,\Omega_0+B]$ as $g(\Omega)$ with $\Omega\in [-1,1]$, where $g(\Omega)$ is predefined as \cite{CommonCodebookDesign}
\begin{equation}\label{codeword obj}
    g(\Omega)=\left\{ \begin{array}{cl}
	\sqrt{2/B}, &\Omega\in\mathcal{I}_{v},\\
	0,&\Omega\notin\mathcal{I}_v.\\
\end{array} \right.
\end{equation}
Then, we can formulate the codeword design problem as
\begin{subequations}\label{CodewordDesignContinue1} \normalsize
\begin{align}
\underset{\boldsymbol{v}}{\min}\ & \frac{1}{2}\int_{-1}^{1}\big(g(\Omega)-\big|\boldsymbol{\alpha}(N_{\rm UE},\Omega)^H\boldsymbol{v}\big|\big)^2 {\rm d} \Omega  \label{Objective1} \\
\mathrm{s.t.\ \ }& \big|[\boldsymbol{v}]_n\big|^2=\frac{1}{N_{\rm UE}}, n=1,2,\cdots,N_{\rm UE} \label{Constraint1}
\end{align}
\end{subequations}
where \eqref{Objective1} aims to minimize the mean-squared error between the predefined beam gain of the codeword and the practical beam gain of $\boldsymbol{v}$, and \eqref{Constraint1} is the constant modulus constraint imposed by the phase shifters of the UEs.

We equally quantize the continuous channel AoA $[-1,1]$ into $Q(Q > N_{\rm UE})$ angles, where the $q$th angle is denoted as
\begin{equation}\label{quantized angle}
  \Omega_q=-1+(2q-1)/Q.
\end{equation}
for $q=1,2,\cdots Q$. Denote
\begin{equation}\label{matrix A}
\boldsymbol{A} \triangleq \sqrt{N_{\rm UE}}[\boldsymbol{\alpha}(N_{\rm UE},\Omega_1),\boldsymbol{\alpha}(N_{\rm UE},\Omega_2),\ldots,\boldsymbol{\alpha}(N_{\rm UE},\Omega_Q)]
\end{equation}
where $\boldsymbol{A}\in\mathbb{C}^{ N_{\rm UE}\times Q}$ is a matrix made up of $Q$ channel steering vectors. Define a vector $\boldsymbol{g}\in \mathbb{C}^{Q}$ with
\begin{equation}\label{vector g}
 [\boldsymbol{g}]_q=g(\Omega_q),~q=1,2,\cdots,Q
\end{equation}
to represent the values of the predefined beam gains along the quantized angles. Then the continuous problem in \eqref{CodewordDesignContinue1} is converted into the discrete problem as
\begin{subequations}\label{CodewordDesignContinue2} \normalsize
\begin{align}
\underset{\boldsymbol{v}}{\min}\ & \big\|\boldsymbol{g}-|\boldsymbol{A}^H\boldsymbol{v}| \big\|_2^2  \label{Objective2} \\
\mathrm{s.t.\ \ }& \big|[\boldsymbol{v}]_n\big|^2=\frac{1}{N_{\rm UE}}, n=1,2,\cdots,N_{\rm UE} \label{Constraint2}.
\end{align}
\end{subequations}
As $Q$ grows to be infinity, the solution of \eqref{CodewordDesignContinue2} will approach the solution of \eqref{CodewordDesignContinue1}. By introducing a phase vector $\boldsymbol{\Theta} \in \mathbb{R}^Q$, we can further rewrite \eqref{CodewordDesignContinue2} as
\begin{subequations}\label{CodewordDesignContinue21} \normalsize
\begin{align}
\underset{\boldsymbol{v},\boldsymbol{\Theta}}{\min}\ & \|\boldsymbol{r}-\boldsymbol{A}^H\boldsymbol{v}\|_2^2  \label{Objective21} \\
\mathrm{s.t.\ \ }& \big|[\boldsymbol{v}]_n\big|^2=\frac{1}{N_{\rm UE}}, n=1,2,\cdots,N_{\rm UE} \label{Constraint21}.
\end{align}
\end{subequations}
where
\begin{equation}\label{DefinitionOfr}
  \boldsymbol{r}\triangleq\boldsymbol{g}\circ e^{j\boldsymbol{\Theta}}
\end{equation}
with $[\boldsymbol{r}]_q = [\boldsymbol{g}]_q e^{j[\boldsymbol{\Theta}]_q}$ for $q=1,2,\ldots,Q$. Note that \eqref{CodewordDesignContinue2} and \eqref{CodewordDesignContinue21} have the same optimal solution of $\boldsymbol{v}$ because we can always design  $\boldsymbol{\Theta}=\angle (\boldsymbol{A}^H\boldsymbol{v})$ so that $\boldsymbol{r}$ and $(\boldsymbol{A}^H\boldsymbol{v})$ have the same phase.

The problem in \eqref{CodewordDesignContinue21} can be solved by the alternative minimization method. We determine $\boldsymbol{\Theta}$ with fixed $\boldsymbol{v}$ and then determine $\boldsymbol{v}$ with fixed $\boldsymbol{\Theta}$, which is repeatedly executed until the maximum number of iterations is reached.

When determining $\boldsymbol{\Theta}$ with fixed $\boldsymbol{v}$, the optimal solution of $\boldsymbol{\Theta}$ can be written as
\begin{equation}\label{OptimalSolutionTheta}
  \boldsymbol{\Theta}=\angle(\boldsymbol{A}^H\boldsymbol{v}).
\end{equation}

When determining $\boldsymbol{v}$ with fixed $\boldsymbol{\Theta}$, \eqref{CodewordDesignContinue21} can be written as
\begin{align}\label{CodewordDesignContinue3}
\underset{\boldsymbol{v}}{\min}\ &\|\boldsymbol{r}-\boldsymbol{A}^H\boldsymbol{v}\|_2^2  \nonumber \\
\mathrm{s.t.\ }&\big|[\boldsymbol{v}]_n\big|^2=\frac{1}{N_{\rm UE}}, n=1,2,\cdots,N_{\rm UE},
\end{align}
which has already been investigated~\cite{TractableBeampattern,WZFan}. In \cite{TractableBeampattern}, the problem is resolved by a successive closed-form (SCF) algorithm, which involves solving a series of convex equality constrained quadratic programs. In \cite{WZFan}, it is resolved by Riemannian optimization algorithm. Different from \cite{TractableBeampattern} and \cite{WZFan}, in this work we will show that the problem has a closed-form expression, which no longer needs running any algorithms and therefore has very low computational complexity.

The objective function in \eqref{CodewordDesignContinue3} can be written as
\begin{align}\label{ObjectiveFunction}
\|\boldsymbol{r}-\boldsymbol{A}^H\boldsymbol{v}\|_2^2&=(\boldsymbol{r}-\boldsymbol{A}^H\boldsymbol{v})^H(\boldsymbol{r}-\boldsymbol{A}^H\boldsymbol{v})\nonumber\\
&=\boldsymbol{r}^H\boldsymbol{r}-\boldsymbol{r}^H\boldsymbol{A}^H\boldsymbol{v}-\boldsymbol{v}^H\boldsymbol{A}\boldsymbol{r}+\boldsymbol{v}^H\boldsymbol{A}\boldsymbol{A}^H\boldsymbol{v}\nonumber\\
&=C-\boldsymbol{p}^H\boldsymbol{v}-\boldsymbol{v}^H\boldsymbol{p}
\end{align}
where $C\triangleq \boldsymbol{r}^H\boldsymbol{r}+\boldsymbol{v}^H\boldsymbol{A}\boldsymbol{A}^H\boldsymbol{v}$ and $\boldsymbol{p} \triangleq \boldsymbol{A}\boldsymbol{r}$. Note that $\boldsymbol{r}^H \boldsymbol{r}=\boldsymbol{g}^H \boldsymbol{g}$ is a constant determined by the predefined absolute beam gain in \eqref{codeword obj}. In addition, we have $\boldsymbol{A}\boldsymbol{A}^H=Q\boldsymbol{I}_{N_{\rm UE}}$ because the entry on the $m$th row and the $n$th column of $\boldsymbol{A}\boldsymbol{A}^H$, for $m=1,2,\cdots N_{\rm UE}$ and $n=1,2,\cdots N_{\rm UE}$, can be expressed as
\begin{align}\label{mth row nth column of AA^H}
[\boldsymbol{A}\boldsymbol{A}^H]_{m,n}&=[\boldsymbol{A}]_{m,:}[\boldsymbol{A}^H]_{:,n}=\sum_{q=1}^{Q} e^{j(m-n)\pi\Omega_q} \nonumber\\
&=\sum_{q=1}^{Q} e^{j(m-n)\pi(-1+\frac{2q-1}{Q})}\nonumber\\
&=e^{j(m-n)\pi(-1-\frac{1}{Q})}\sum_{q=1}^{Q}e^{j(m-n)\pi\frac{2q}{Q}}\nonumber\\
&=\left\{ \begin{array}{cl}
	Q , &m=n,\\
	0,& {\rm others}.\\
\end{array} \right.
\end{align}
Then we have $\boldsymbol{v}^H\boldsymbol{A}\boldsymbol{A}^H\boldsymbol{v}=Q\boldsymbol{v}^H\boldsymbol{v}$, which is also a constant due to the constant modulus constraint of $\boldsymbol{v}$. Therefore, $C$ is a constant and \eqref{CodewordDesignContinue3} can be convert into
\begin{align}\label{CodewordDesignContinue4}
\underset{\boldsymbol{v}}{\max}\ & \boldsymbol{p}^H\boldsymbol{v}+\boldsymbol{v}^H\boldsymbol{p}  \nonumber \\
\mathrm{s.t.\ \ }&\big|[\boldsymbol{v}]_n\big|^2=\frac{1}{N_{\rm UE}}, n=1,2,\cdots,N_{\rm UE}.
\end{align}
Obviously, \eqref{CodewordDesignContinue4} is equivalent to the following problem
\begin{align}\label{CodewordDesignContinue5}
\underset{\boldsymbol{u}}{\max}\ & \boldsymbol{t}^T\boldsymbol{u}  \nonumber \\
\mathrm{s.t.\ \ }& [\boldsymbol{u}]_n^2+[\boldsymbol{u}]_{n+N_{\rm UE}}^2=\frac{1}{N_{\rm UE}}, n=1,2,\cdots,N_{\rm UE}
\end{align}
where
\begin{equation}\label{subequation of CodewordDesignContinue5}
\boldsymbol{t}=\begin{bmatrix} {\rm Re}\{\boldsymbol{p}\} \\ {\rm Im}\{\boldsymbol{p}\} \end{bmatrix},~\boldsymbol{u}=\begin{bmatrix} {\rm Re}\{\boldsymbol{v}\} \\ {\rm Im}\{\boldsymbol{v}\} \end{bmatrix}.
\end{equation}

Note that \eqref{CodewordDesignContinue5} can be divided into $N_{\rm UE}$ mutually independent subproblems, where the $n$th subproblem for $n=1,2,\cdots,N_{\rm UE}$ can be written as
\begin{align}\label{SubCodewordDesignContinue}
\underset{[\boldsymbol{u}]_n,[\boldsymbol{u}]_{n+N_{\rm UE}}}{\max}\ & [\boldsymbol{t}]_n[\boldsymbol{u}]_n+[\boldsymbol{t}]_{n+N_{\rm UE}}[\boldsymbol{u}]_{n+N_{\rm UE}}  \nonumber \\
\mathrm{s.t.\ \ \ \ \ \ }& [\boldsymbol{u}]_n^2+[\boldsymbol{u}]_{n+N_{\rm UE}}^2=\frac{1}{N_{\rm UE}}, n=1,2,\cdots,N_{\rm UE}.
\end{align}
The optimal solution of \eqref{SubCodewordDesignContinue} can be easily computed as
\begin{align}\label{optimalsolution1}
&[\boldsymbol{u}]_n=\frac{[\boldsymbol{t}]_n}{\sqrt{N_{\rm UE}([\boldsymbol{t}]_n^2+[\boldsymbol{t}]_{n+N_{\rm UE}}^2)}},\nonumber\\&[\boldsymbol{u}]_{n+N_{\rm UE}}=\frac{[\boldsymbol{t}]_{n+N_{\rm UE}}}{\sqrt{N_{\rm UE}([\boldsymbol{t}]_n^2+[\boldsymbol{t}]_{n+N_{\rm UE}}^2)}}.
\end{align}
According to \eqref{subequation of CodewordDesignContinue5}, we have
\begin{equation}\label{optimalsolution2}
[\boldsymbol{v}]_n=[\boldsymbol{u}]_n+j[\boldsymbol{u}]_{n+N_{\rm UE}},~n=1,2,\cdots,N_{\rm UE}
\end{equation}
which is the closed-form expression for \eqref{CodewordDesignContinue3}. Therefore, we can determine $\boldsymbol{v}$ with fixed $\boldsymbol{\Theta}$.

\begin{algorithm}[!t]
	\caption{AMCF-ZCI Codeword Design for the UE}
	\label{alg-ACF}
	\begin{algorithmic}[1]
        \STATE \textbf{Input:} $N_{\rm UE}$, $\Omega_0$, $B$ and $M$.
        \STATE Obtain $\boldsymbol{v}^{(0)}$ via \eqref{Zadoff-Chu Initial}.
        \STATE Set $m=1$.
        \WHILE{$m \leq M$}
        \STATE Obtain $\boldsymbol{\Theta}^{(m)}=\angle (\boldsymbol{A}^H\boldsymbol{v}^{(m-1)})$ according to \eqref{OptimalSolutionTheta}.
        \STATE Obtain $\boldsymbol{r}^{(m)}=\boldsymbol{g} \circ e^{j\boldsymbol{\Theta}^{(m)}}$ according to \eqref{DefinitionOfr}.
        \STATE Obtain $\boldsymbol{v}^{(m)}$ via \eqref{optimalsolution2}.
        \STATE $m\leftarrow m+1$.
        \ENDWHILE
        \STATE \textbf{Output:} $\boldsymbol{v}_{\rm o}=\boldsymbol{v}^{(M)}$.
	\end{algorithmic}
\end{algorithm}

We alternatively optimize $\boldsymbol{v}$ and $\boldsymbol{\Theta}$ until a predefined maximum number of iterations $M$ is reached. To speed up the convergence of the AMCF, we consider the following two different initializations.

\subsection{Zadoff-Chu Sequence Initialization}
According to~\cite{RuiPeng}, we can initialize the $n$th entry of $\boldsymbol{v}$ as
\begin{equation}\label{Zadoff-Chu Initial}
[\boldsymbol{v}^{(0)}]_n=\left\{ \begin{array}{ll}
	\frac{1}{\sqrt{N_{\rm UE}}}e^{j\pi (\frac{B n^2}{2N_{\rm UE}}+n\Omega_0)}, &~N_{\rm UE} {\rm~is~even}\\
    \frac{1}{\sqrt{N_{\rm UE}}}e^{j\pi (\frac{B n(n+1)}{2N_{\rm UE}}+n\Omega_0)},&~N_{\rm UE} {\rm~is~odd}\\
\end{array} \right.
\end{equation}
for $n=1,2,\ldots, N_{\rm UE}$, where $\boldsymbol{v}^{(0)}$ is essentially a variant of Zadoff-Chu sequence in \cite{Zadoff-Chu}.

Although the beam generated by $\boldsymbol{v}^{(0)}$ covers the angle space of $[\Omega_0,\Omega_0+B]$, $\boldsymbol{v}^{(0)}$ cannot be taken as a good codeword for beam training. To be specific, in the transition zone of the beam generated by $\boldsymbol{{v}}^{(0)}$, there are lots of fluctuations, which can deteriorate the performance of beam training. To tackle this issue, additional hardware such as a group of unequal power dividers, is used in the phase shifter network so that the power of each antenna can be changed to smooth the transition zone~\cite{RuiPeng}.

Different from~\cite{RuiPeng} that uses a group of unequal power dividers, in this work we avoid any additional hardware. We just use $\boldsymbol{v}^{(0)}$ as the initialization of AMCF and use the output of AMCF $\boldsymbol{v}_{\rm o}$ as a designed codeword for the UE.

The steps of the AMCF with Zadoff-Chu sequence initialization (AMCF-ZCI) are summarized in \textbf{Algorithm 1}.

Finally, based on the aforementioned codeword design schemes, the  hierarchical codebook for each UE can be designed as follows.
\begin{enumerate}
\item Initialize the layer counter of the codebook as $s=1$ and the left boundary of the beam coverage as $\Omega_0=-1$. Then set $B=2/2^s$.
\item Design $\boldsymbol{\mathcal{V}}_{\rm UE}(s,1)$ by \textbf{Algorithm 1}.
\item Obtain $\boldsymbol{\mathcal{V}}_{\rm UE}(s,m)$, for $m=2,\cdots 2^s$, based on the following equation
\begin{equation}\label{CodewordShifted}
\boldsymbol{\mathcal{V}}_{\rm UE}(s,m)=\sqrt{N_{\rm UE}} \boldsymbol{\mathcal{V}}_{\rm UE}(s,1)\circ \boldsymbol{\alpha}(N_{\rm UE},(m-1)/2^{s-1})
\end{equation}
which is essentially the shifted version of $\boldsymbol{\mathcal{V}}_{\rm UE}(s,1)$ for different beam coverage.
\item  Increase $s$ by one, i.e., $s\leftarrow s+1$.
\item Repeat the steps 2) to 4) until reaching the last layer of $\boldsymbol{\mathcal{V}}_{\rm UE}$, i.e., $s=T+1$.
\end{enumerate}

\section{Simultaneous Beam Training based on Adaptive Hierarchical Codebook}\label{Sec.BS}
In this section, we will propose a simultaneous multiuser beam training scheme based on adaptive hierarchical codebook, which considerably reduces the training overhead compared to the existing hierarchical beam training.

\subsection{Simultaneous Multiuser Hierarchical Beam Training}
We denote the hierarchical codebook for the BS as $\boldsymbol{\mathcal{C}}$ to distinguish with the existing hierarchical codebook $\boldsymbol{\mathcal{V}}_{\rm BS}$.

\begin{figure}[!t]
\centering
\includegraphics[width=84mm]{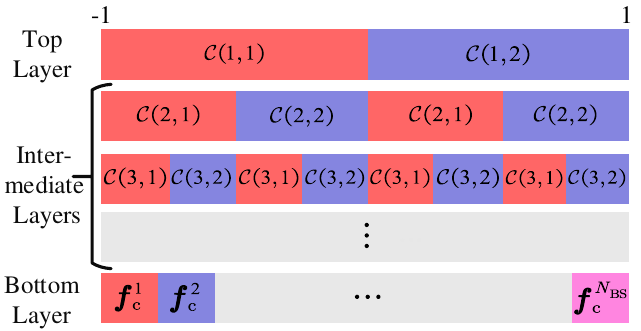}
\caption{Illustration of adaptive hierarchical codebook $\boldsymbol{\mathcal{C}}$.}
\label{CodebookNew}
\end{figure}

As shown in Fig.~\ref{CodebookNew}, the adaptive hierarchical codebook with totally $S=\log_2 {N_{\rm BS}}$ layers can be divided into the top layer, the bottom layer and the intermediate layers.
\begin{enumerate}
\item In the top layer of the codebook, we  equally  divide the channel AoD $[-1,1]$ into two codewords, so that the beam width of each codeword is one.
\item The bottom layer that is the $S$th layer of the hierarchical codebook, is exactly the same as the bottom layer of the existing hierarchical codebook and can be designed according to \eqref{Codebook}.
\item The intermediate layers consist of the second to the  $(S-1)$th layers of the codebook. Different from the existing codebook, each intermediate layer only includes two codewords, no matter how large $K$ is. The beam coverage of codewords in the intermediate layers is intermittent. In particular, the codewords are adaptively designed according to the estimated channel AoDs in the previous layer. Note that the summation of the beam coverage of two codewords in the same layer may not be $[-1,1]$ because some regions may not have any channel path and we do not need to waste signal beam to cover them.
\end{enumerate}

Now we focus on designing codewords in the first and intermediate layers of the adaptive hierarchical codebook. In general, the beam coverage of a codeword at the upper layer can be considered as the union of several codewords at the bottom layer. Therefore, we can design the codewords in the first and intermediate layers by combining several codewords from the bottom layer of $\boldsymbol{\mathcal{C}}$. Then the $m$th codeword at the $s$th layer of $\boldsymbol{\mathcal{C}}$, denoted as $\boldsymbol{\mathcal{C}}(s,m)$, for $s=1,\cdots,S-1$ and $m=1,2,\cdots2^s$, can be represented as
\begin{equation}\label{Codeword}
\boldsymbol{\mathcal{C}}(s,m)=\sum_{n\in\boldsymbol{\Psi}_{s,m}}e^{j\psi_n}\boldsymbol{f}_{\rm c}^n
\end{equation}
which is essentially a weighted summation of several channel steering vectors. The indices of the codewords of $\boldsymbol{\mathcal{F}}$ involved in the weighted summation form an integer set $\boldsymbol{\Psi}_{s,m}$. Here we introduce phase $\psi_n$ to explore the additional degree of freedom to avoid low beam gain within the beam coverage~\cite{PS-DFT2017}. Based on our previous work \cite{CKJLETTER,TSPCKJ}, we can set
\begin{equation}\label{Phi}
  \psi_n=n\pi\bigg(-1+\frac{1}{N_{\rm BS}}\bigg).
\end{equation}
To fairly compare different codewords in each test, we usually normalize $\boldsymbol{\mathcal{C}}(s,m)$ so that $\|\boldsymbol{\mathcal{C}}(s,m)\|_2=1$.

We design $\boldsymbol{\Psi}_{s,m}$, for $s=1,\cdots,S-1$ and $m=1,2,\cdots2^s$, as follows. We denote the beam coverage of $\boldsymbol{\mathcal{C}}(s,m)$ as $B_{s,m}$. Then $B_{1,m}$ at the top layer can be expressed as
\begin{equation}\label{first part}
B_{1,m}=[m-2,m-1],~m=1,2.
\end{equation}
We determine $\boldsymbol{\Psi}_{1,m}$ by
\begin{align}\label{index set}
\boldsymbol{\Psi}_{1,m}=\Bigg\{n\Bigg |\bigg[-1+\frac{2n-2}{N_{\rm BS}},-&1+\frac{2n}{N_{\rm BS}}\bigg] \subseteq B_{1,m}, \nonumber \\
&n=1,2,\cdots,N_{\rm BS}\Bigg\}
\end{align}
for $m=1,2$, where $[-1+\frac{2n-2}{N_{\rm BS}},-1+\frac{2n}{N_{\rm BS}}]$ denotes the beam coverage of $\boldsymbol{f}_{\rm c}^n$.



The BS sequentially transmits $\boldsymbol{\mathcal{C}}(1,1)$ and $\boldsymbol{\mathcal{C}}(1,2)$ to all $K$ UEs and each UE receives the signal with $\boldsymbol{\mathcal{V}}_{\rm UE}(1,1)$ and $\boldsymbol{\mathcal{V}}_{\rm UE}(1,2)$, respectively. Then each UE compares the received signal power of $\boldsymbol{\mathcal{C}}(1,1)$ and $\boldsymbol{\mathcal{C}}(1,2)$ and individually feeds back the index of the larger codeword to the BS. Denote $\mathcal{K} \triangleq \{1,2,\ldots,K\}$. We define a vector $\boldsymbol{\Gamma}_1$ of length of $K$, where the $k(k\in \mathcal{K})$th entry of $\boldsymbol{\Gamma}_1$ denoted as $[\boldsymbol{\Gamma}_1]_k$ corresponds to the index of the larger codewords from the $k$th UE, i.e., $[\boldsymbol{\Gamma}_1]_k\in\{1,2\}$.


Denote the index set of the selected codewords after beam training at the $(s-1)$th layer to be  $\boldsymbol{\Gamma}_{s-1}$ for $s=2,3,\cdots,S-1$. According to the existing hierarchical beam training, at the $s$th layer, the BS will test $\boldsymbol{\mathcal{V}}_{\rm BS}(s, 2[\boldsymbol{\Gamma}_{s-1}]_k-1)$ and $\boldsymbol{\mathcal{V}}_{\rm BS}(s, 2[\boldsymbol{\Gamma}_{s-1}]_k)$, which are the refined codewords of $\boldsymbol{\mathcal{V}}_{\rm BS}(s-1, [\boldsymbol{\Gamma}_{s-1}]_k)$. Therefore, it needs totally $2K$ times of beam training to test all $K$ UEs. To reduce the training overhead, we consider the following two cases:
\begin{enumerate}
\item If $[\boldsymbol{\Gamma}_{s-1}]_i=[\boldsymbol{\Gamma}_{s-1}]_q~(i \in \mathcal{K},q\in \mathcal{K}, i\neq q)$ that is the $i$th UE and the $q$th UE share the same AoD at the $(s-1)$th layer of $\boldsymbol{\mathcal{V}}_{\rm BS}$, we can perform beam training for them simultaneously because they will have the same refined codewords at the $s$th layer of $\boldsymbol{\mathcal{V}}_{\rm BS}$.
 \item   If $[\boldsymbol{\Gamma}_{s-1}]_i \ne [\boldsymbol{\Gamma}_{s-1}]_q~(i \in \mathcal{K},q\in \mathcal{K}, i\neq q)$, which means that the $i$th UE and the $q$th UE have different AoD at the $(s-1)$th layer of $\boldsymbol{\mathcal{V}}_{\rm BS}$, the $i$th UE cannot receive the signal transmitted from the beam coverage of $\boldsymbol{\mathcal{V}}_{\rm BS}(s-1, [\boldsymbol{\Gamma}_{s-1}]_q)$ because the AoD of the  $i$th UE is located in the beam coverage of $\boldsymbol{\mathcal{V}}_{\rm BS}(s-1, [\boldsymbol{\Gamma}_{s-1}]_i)$. Therefore, we can distinguish different UEs based on their different AoDs. We will show in the following texts that the BS can also simultaneously perform the beam training for these two UEs.
 \end{enumerate}

 \begin{figure}[!t]
\centering
\includegraphics[width=80mm]{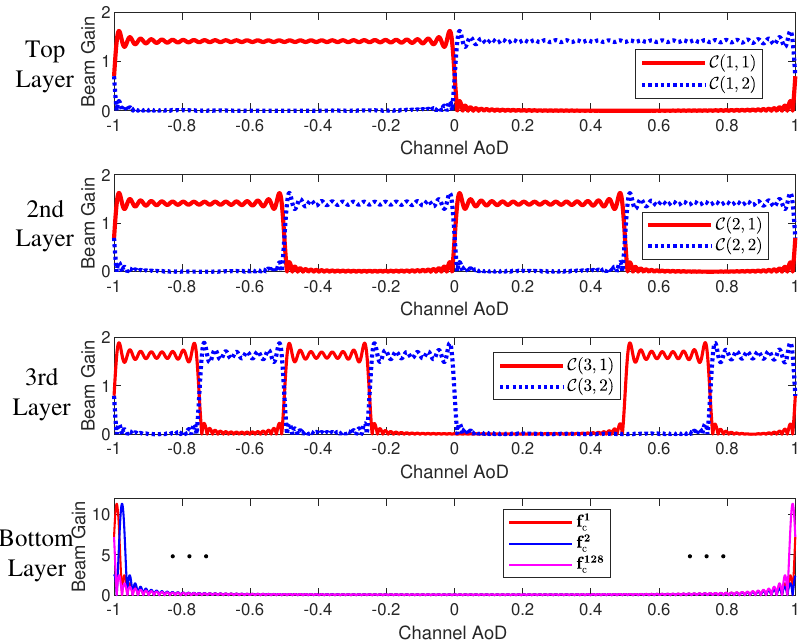}
\caption{Beam gain of different codewords in adaptive hierarchical codebook $\boldsymbol{\mathcal{C}}$ with $N_{\rm BS}=128$, $N_{\rm UE}=16$ and $K=4$.}
\label{BeamIllustration}
\end{figure}

Based on the above discussion, in either case, we suppose $\boldsymbol\Gamma_{s-1}$ has $K'(K' \leq K)$ different integers, which corresponds to $K'$ different codewords at the $(s-1)$th layer and $2K'$ refined codewords at the $s$th layer of $\boldsymbol{\mathcal{V}}_{\rm BS}$. In the proposed simultaneous multiuser beam training scheme, we divide these $2K'$ refined codewords into two groups. The union of beam coverage of $K'$ codewords in the first and second group are respectively denoted as
\begin{equation}\label{beam coverage union1}
 \left\{\begin{array}{ll}
B_{s,1}=\bigcup_mD_{s,m},~{\rm if}~\frac{m+1}{2}\in \boldsymbol{\Gamma}_{s-1},~m=1,2,\cdots,2^s,\\
B_{s,2}=\bigcup_mD_{s,m},~{\rm if}~\frac{m}{2}\in \boldsymbol{\Gamma}_{s-1},~m=1,2,\cdots,2^s
\end{array}\right.
\end{equation}
where $D_{s,m}=[-1+(m-1)/2^{s-1},-1+m/2^{s-1}]$ is the beam coverage of $\boldsymbol{\mathcal{V}}_{\rm BS}(s,m)$. It is seen that the beam coverage of $B_{s,1}$ and $ B_{s,2}$ is intermittent, which is the motivation of our work to design multi-mainlobe codewords. Then we can obtain $\boldsymbol{\Psi}_{s,1}$ and $\boldsymbol{\Psi}_{s,2}$ based on $ B_{s,1}$ and $ B_{s,2}$ respectively via
\begin{align}\label{Phi_com}
\boldsymbol{\Psi}_{s,m}=\bigg\{n\bigg |\bigg[-1+\frac{2n-2}{N_{\rm BS}},-&1+\frac{2n}{N_{\rm BS}}\bigg]\subset B_{s,m}, \nonumber\\
&n=1,2,\cdots,N_{\rm BS}\bigg\}
\end{align}
for $m=1,2$. Given $\boldsymbol{\Psi}_{s,1}$ and $\boldsymbol{\Psi}_{s,2}$, we can design $\boldsymbol{\mathcal{C}}(s,1)$ and $\boldsymbol{\mathcal{C}}(s,2)$, respectively via \eqref{Codeword}. Therefore, by using \eqref{beam coverage union1}, \eqref{Phi_com} and \eqref{Codeword}, we can design $\boldsymbol{\mathcal{C}}(s,1)$ and $\boldsymbol{\mathcal{C}}(s,2)$ based on the beam training results of the $(s-1)$th layer, i.e., $\boldsymbol\Gamma_{s-1}$. Note that either $\boldsymbol{\mathcal{C}}(s,1)$ or $\boldsymbol{\mathcal{C}}(s,2)$ is a multi-mainlobe codeword, where each mainlobe covers a spatial region that one or more users are probably in. When the number of UEs increases, the number of mainlobes of the multi-mainlobe codeword may also grow; but the number of codewords keeps to be two in each layer excluding the bottom layer of the hierarchical codebook.

At the $s$th layer of $\boldsymbol{\mathcal{C}}$, for $s=2,3,\ldots,S-1$, the BS sequentially transmits $\boldsymbol{\mathcal{C}}(s,1)$ and $\boldsymbol{\mathcal{C}}(s,2)$ to all $K$ UEs. Note
that there are only two codewords at each intermediate layer,
which only requires two times of simultaneous beam training
for all the UEs no matter how many UEs the BS serves. Then each UE compares the received signal power of $\boldsymbol{\mathcal{C}}(s,1)$ and $\boldsymbol{\mathcal{C}}(s,2)$ and individually feeds back the index of the larger codeword to the BS. We define $\boldsymbol{\Phi}_s$ as a vector of length of $K$ to keep the indices fed back by all $K$ UEs, where $[\boldsymbol{\Phi}_s]_k$ is the index fed back by the $k$th UE. In fact, $[\boldsymbol{\Phi}_s]_k\in \{1,2\}$. Then we can obtain $\boldsymbol{\Gamma}_s$ by
\begin{equation}\label{Phi_s}
[\boldsymbol{\Gamma}_s]_k=2\big([\boldsymbol{\Gamma}_{s-1}]_k-1\big)+[\boldsymbol{\Phi}_s]_k,
\end{equation}
which can be used to determine $B_{s+1,1}$ and $B_{s+1,2}$ via \eqref{beam coverage union1} for $k=1,2,\cdots,K$. We iteratively perform these steps until arriving at the bottom layer of $\boldsymbol{\mathcal{C}}$.

At the bottom layer of $\boldsymbol{\mathcal{C}}$, different from the downlink beam training in the top and intermediate layers, we perform the uplink beam training so that each entry of the effective channel matrix in \eqref{effective} can be obtained. During the uplink beam training between the $k$th UE and the BS, the BS sequentially uses $\boldsymbol{f}_{\rm c}^{2[\boldsymbol{\Gamma}_{S-1}]_k-1}$ and $\boldsymbol{f}_{\rm c}^{2[\boldsymbol{\Gamma}_{S-1}]_k}$ to receive the signal and selected one with the larger received signal power. In this way, the best BS codeword in $\boldsymbol{\mathcal{F}}$ for the $k$th UE can be determined. Note that the BS has $N_{\rm RF}$ RF chains, which implies that the BS can use multiple RF chains for parallel signal receiving to improve the efficiency~\cite{TWCSXY,TSPHeShiwen}. Therefore, totally $2K$ times of beam training are required at the bottom layer of $\boldsymbol{\mathcal{C}}$.
Similar to~\eqref{Phi_s}, we can obtain $\boldsymbol{\Gamma}_S$ by
\begin{equation}\label{Phi_T}
[\boldsymbol{\Gamma}_S]_k=2\big([\boldsymbol{\Gamma}_{S-1}]_k-1\big)+[\boldsymbol{\Phi}_S]_k,~k=1,2,\cdots,K.
\end{equation}

Finally, the $k$th column of the designed analog precoder $\boldsymbol{\widehat{F}}_{\rm RF}$, denoted as $\boldsymbol{\widehat{f}}_k$, can be obtained via
\begin{equation}\label{FRFcolumnwise}
  \boldsymbol{\widehat{f}}_k=\boldsymbol{f}_{\rm c}^{[\boldsymbol{\Gamma}_S]_k}.
\end{equation}

The detailed steps of the proposed simultaneous multiuser hierarchical beam training is summarized in \textbf{Algorithm} \ref{alg-SHBT}.

\begin{table*}[!t]
\centering
\caption{Comparisons of overhead for different schemes.}
\label{Tab.Complexity}
\begin{tabular}{ p{6cm}p{5cm}p{4cm}}
\toprule
Schemes  &   Training Overhead &   Feedback Overhead\\
\midrule
Our scheme &  $2(K+\log_2 N_{\rm UE} N_{\rm BS}-1)$   & $K(\log_2 N_{\rm BS}-1)$ \\
Scheme in~\cite{AlMultiUser}  & $N_{\rm BS}N_{\rm UE}$ & $K$ \\
TDMA hierarchical beam training  &   $2K(\log_2 N_{\rm BS}+\log_2 N_{\rm UE})$ &  $K\log_2 N_{\rm BS}$\\
\bottomrule
\end{tabular}
\end{table*}
\begin{algorithm}[!t]
	\caption{Simultaneous Multiuser Hierarchical Beam Training}
	\label{alg-SHBT}
	\begin{algorithmic}[1]
        \STATE \textbf{Input:} $N_{\rm BS}$, $N_{\rm UE}$ and $K$.
        \STATE Obtain $\boldsymbol{\mathcal{C}}(1,1)$ and $\boldsymbol{\mathcal{C}}(1,2)$ via \eqref{index set} and \eqref{Codeword}.
        \STATE Obtain $\boldsymbol{\Gamma}_1$ by the top layer beam training.
        \STATE Set $S=\log_2 N_{\rm BS}$.
        \FOR {$s=2,3,\ldots,S-1$}
        \STATE Obtain $B_{s,1}$ and $B_{s,2}$ via \eqref{beam coverage union1}.
        \STATE Obtain $\boldsymbol{\Psi}_{s,1}$ and $\boldsymbol{\Psi}_{s,2}$ via \eqref{Phi_com}.
        \STATE Generate $\boldsymbol{\mathcal{C}}(s,1)$ and $\boldsymbol{\mathcal{C}}(s,2)$  via  \eqref{Codeword}.
        \STATE Obtain $\boldsymbol{\Gamma}_s$ via \eqref{Phi_s}.
         \ENDFOR
         \STATE Obtain  $\boldsymbol{\Gamma}_S$ via \eqref{Phi_T}.
         \STATE Obtain $\boldsymbol{\widehat{f}}_k$ via \eqref{FRFcolumnwise}.
        \STATE \textbf{Output:} $\{\boldsymbol{\widehat{f}}_k, k=1,2,\ldots,K\}$.
	\end{algorithmic}
\end{algorithm}

When designing codewords in the top and intermediate layers of $\boldsymbol{\mathcal{C}}$, we first obtain the ideal codewords by the weighted summation of channel steering vectors as \eqref{Codeword} and then obtain the practical codewords regarding the number of RF chains and the resolution of phase shifters to approximate the ideal codewords based on the method in \cite{TSPCKJ}.

Now we give an example for the proposed simultaneous multiuser hierarchical beam training with $N_{\rm BS}=128$, $N_{\rm UE}=16$ and $K=4$. As shown in Fig.~\ref{BeamIllustration}, we illustrate the beam gain of different codewords in $\boldsymbol{\mathcal{C}}$. To improve the readability of our scheme, each layer in Fig.~\ref{BeamIllustration} corresponds to that in Fig.~\ref{CodebookNew}. At the top layer of $\boldsymbol{\mathcal{C}}$, the BS sequentially transmits $\boldsymbol{\mathcal{C}}(1,1)$ and $\boldsymbol{\mathcal{C}}(1,2)$. The union of beam coverage of $\boldsymbol{\mathcal{C}}(1,1)$ and $\boldsymbol{\mathcal{C}}(1,2)$ equals the full space of $[-1,1]$, since the BS has no knowlege of the UEs. After the top layer beam training, we suppose the indices fed back from the four UEs form a set $\boldsymbol{\Gamma}_1=\{1,1,2,2\}$, which indicates that the channel AoDs of the first and second UEs happen to locate in the same beam coverage of $\boldsymbol{\mathcal{C}}(1,1)$, and the channel AoDs of the third and fourth UEs happen to locate in the same beam coverage of $\boldsymbol{\mathcal{C}}(1,2)$. Based on $\boldsymbol{\Gamma}_1$, we can obtain $\boldsymbol{\Psi}_{2,1}=\{1,2\cdots,32,65,66,\cdots,96\}$ and $\boldsymbol{\Psi}_{2,2}=\{33,34\cdots,64,97,98,\cdots,128\}$ via \eqref{Phi_com}. Based on $\boldsymbol{\Psi}_{2,1}$ and $\boldsymbol{\Psi}_{2,2}$, we can design two multi-mainlobe codewords $\boldsymbol{\mathcal{C}}(2,1)$ and $\boldsymbol{\mathcal{C}}(2,2)$ via \eqref{Codeword}. During the second layer of beam training, the BS sequentially transmits $\boldsymbol{\mathcal{C}}(2,1)$ and $\boldsymbol{\mathcal{C}}(2,2)$. Suppose the indices fed back from the four UEs form a set $\boldsymbol{\Phi}_2=\{1,2,2,2\}$, where each entry denotes the codeword index $\{i|\boldsymbol{\mathcal{C}}(2,i),i=1,2\}$ with the larger received signal power at the UEs. Then we can obtain $\boldsymbol{\Gamma}_2=\{1,2,4,4\}$ via \eqref{Phi_s}. Based on $\boldsymbol{\Gamma}_2$, we can design $\boldsymbol{\Psi}_{3,1}=\{1,2,\cdots,16,33,34,\cdots,48,97,98\cdots,112\}$ and $\boldsymbol{\Psi}_{3,2}=\{17,18,\cdots,32,49,50,\cdots,64,113,114,\cdots,128\}$ via \eqref{Phi_com}. Based on $\boldsymbol{\Psi}_{3,1}$ and $\boldsymbol{\Psi}_{3,2}$, we can design two multi-mainlobe codewords $\boldsymbol{\mathcal{C}}(3,1)$ and $\boldsymbol{\mathcal{C}}(3,2)$ via \eqref{Codeword}. Note that both $\boldsymbol{\mathcal{C}}(3,1)$ and $\boldsymbol{\mathcal{C}}(3,2)$ have three mainlobes. We repeat these procedures until arriving at the bottom layer of $\boldsymbol{\mathcal{C}}$.

\subsection{Digital Precoding}
Note that the beam training and data transmission are two different stages of mmWave massive MIMO communications, where the former is to obtain the CSI that will be used for the latter. During the beam training, the BS employs a hierarchical codebook with multi-mainlobe codewords to serve all the users, where multiple RF chains might be used to generate multi-mainlobe codewords. Once the beam training is finished, the BS finds a best codeword $\boldsymbol{\widehat{f}}_k$ for the $k$th user, where $\boldsymbol{\widehat{f}}_k$ can be generated by a single RF chain according to~\eqref{Codebook}. Since the designed analog combiner, $\boldsymbol{\widehat{w}}_k$, for the $k$th UE can be obtained by the existing hierarchical beam training method, the details are omitted in this work due to the page limitation.

In the following, we design the digital precoding for the data transmission. Stacking $\{y_k, k=1,2,\ldots,K\}$ in \eqref{Received Signal} together as $\boldsymbol{y}=[y_1,y_2,\ldots,y_K]^T$, we have
\begin{equation}\label{received signal}
  \boldsymbol{y}=\boldsymbol{H}_{\rm e}\boldsymbol{F}_{\rm BB}\boldsymbol{s},
\end{equation}
where
\begin{equation}\label{effective}
  \boldsymbol{H}_{\rm e}=\left[\begin{matrix}
   \boldsymbol{\widehat{w}}_1^H\boldsymbol{H}_1\boldsymbol{\widehat{f}}_1\ & \boldsymbol{\widehat{w}}_1^H\boldsymbol{H}_1\boldsymbol{\widehat{f}}_2 & \cdots & \boldsymbol{\widehat{w}}_1^H\boldsymbol{H}_1\boldsymbol{\widehat{f}}_K \\
   \boldsymbol{\widehat{w}}_2^H\boldsymbol{H}_2\boldsymbol{\widehat{f}}_1\ & \boldsymbol{\widehat{w}}_2^H\boldsymbol{H}_2\boldsymbol{\widehat{f}}_2 & \cdots & \boldsymbol{\widehat{w}}_2^H\boldsymbol{H}_2\boldsymbol{\widehat{f}}_K \\
   \vdots & \cdots & \ddots &\vdots \\
   \boldsymbol{\widehat{w}}_K^H\boldsymbol{H}_K\boldsymbol{\widehat{f}}_1\ & \boldsymbol{\widehat{w}}_K^H\boldsymbol{H}_K\boldsymbol{\widehat{f}}_2 & \cdots & \boldsymbol{\widehat{w}}_K^H\boldsymbol{H}_K\boldsymbol{\widehat{f}}_K \\
  \end{matrix}\right]
\end{equation}
is defined as the effective channel matrix. Note that each entry of $\boldsymbol{H}_{\rm e}$ can be obtained via the uplink beam training at the bottom layer of $\boldsymbol{\mathcal{C}}$. If two users are geographically close to each other, they may share the same BS codeword, e.g., $\boldsymbol{\widehat{f}}_1=\boldsymbol{\widehat{f}}_2$, which will cause the rank deficiency of $\boldsymbol{H}_{\rm e}$. Since the digital precoding requires $\boldsymbol{H}_{\rm e}$ to be full rank, we need to make beam allocation for different users to avoid beam conflict, which has already been addressed in~\cite{TWCSXY} and is out of scope of this paper.

The designed digital precoder under the zero forcing (ZF) criterion can be expressed as
\begin{equation}\label{ZF precoder}
\boldsymbol{\widehat{F}}_{\rm BB} = \boldsymbol{{H}}_{\rm e}^H(\boldsymbol{{H}}_{\rm e}\boldsymbol{{H}}_{\rm e}^H)^{-1}.
\end{equation}

\subsection{Overhead Analysis}
At the top layer of $\boldsymbol{\mathcal{C}}$, the BS sequentially transmits two codewords and each UE receives the signal with two codewords, which occupies $4$ time slots. At the intermediate layers of $\boldsymbol{\mathcal{C}}$ from $s=2$ to $s=\log_2 N_{\rm UE}$, the BS sequentially transmits two codewords and each UE receives signal with two codewords, which occupies totally $4(\log_2 N_{\rm UE}-1)$ time slots. At the intermediate layers of $\boldsymbol{\mathcal{C}}$ from $s=\log_2 N_{\rm UE}+1$ to $s=\log_2 N_{\rm BS}-1$, where the hierarchical beam training at the UEs has already been finished, the BS transmits two codewords and each UE receives signal with a single codeword, which occupies totally $2(\log_2 N_{\rm BS}-\log_2 N_{\rm UE}-1)$ time slots. At the bottom layer of $\boldsymbol{\mathcal{C}}$, the BS uses two codewords to receive the signal from each UE, which results in totally $2K$ time slots. In all, our proposed scheme needs totally $(2K+2\log_2 N_{\rm UE}N_{\rm BS}-2)$ time slots. As shown in  Table.~\ref{Tab.Complexity}, we compare the training overhead of different schemes. For example, if $N_{\rm BS}=128$, $N_{\rm UE}=16$, $K=8$, our scheme, the scheme in~\cite{AlMultiUser} and the TDMA hierarchical beam training require $36$, $2048$ and $176$ time slots, respectively. Compared to the latter two schemes, our scheme can reduce the training overhead by $98.2\%$ and $79.6\%$, respectively.

We also compare the feedback overhead from the UEs to the BS. For the scheme in~\cite{AlMultiUser}, each UE needs only one time of feedback after finishing the beam sweeping, which results in totally $K$ times of feedback. Since we do need feedback at the bottom layer in our scheme, the feedback times of our scheme is $K$ less than that of the TDMA hierarchical beam training.

\section{Simulation Results}\label{Sec.Simulation}
Now we evaluate our schemes for multiuser mmWave massive MIMO systems by simulation.

\subsection{Evaluation of codeword design schemes for each UE}
To evaluate the codeword design schemes for each UE, we consider a single-user mmWave massive MIMO system, where the BS equipped with $N_{\rm BS}=128$ antennas serves only one UE for simplification, since there is no difference in codebook design between one UE and multiple UEs. The UE is supposed to be equipped with $N_{\rm UE}=32$ antennas and a single RF chain.

\begin{figure}[!t]
\centering
\includegraphics[width=90mm]{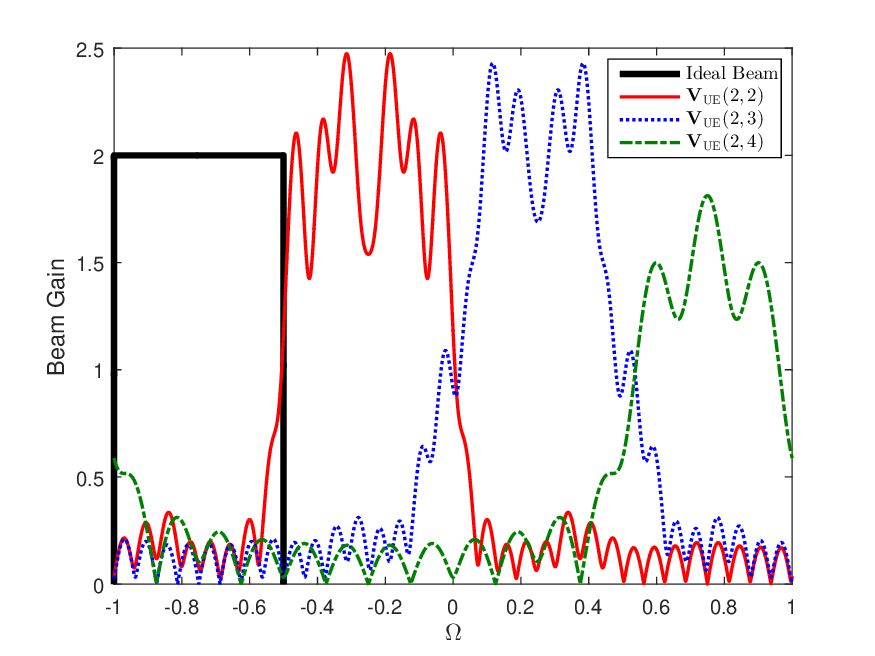}
\caption{Comparisons of beam patterns using different codeword design schemes for each UE.}
\label{BeamPatternCompare}
\end{figure}

Fig.~\ref{BeamPatternCompare} compares the beam patterns for each UE using different codeword design schemes. For fair comparison, the beam width is set as $B=1/2$, which is typically set for the codewords at the second layer of $\boldsymbol{V}_{\rm UE}$. Given $N_{\rm UE}=32$, the constant modulus constraint on the antenna weights is $1/\sqrt{32}$. According to \eqref{codeword obj}, the ideal beam gain is
\begin{equation}\label{codeword obj2}
    g(\Omega)=\left\{ \begin{array}{cl}
	2, &\Omega\in\mathcal{I}_{v},\\
	0,&\Omega\notin\mathcal{I}_v,\\
\end{array} \right.
\end{equation}
which can form the beam pattern illustrated by the black solid line in Fig.~\ref{BeamPatternCompare}.  The beam patterns of $\boldsymbol{\mathcal{V}}_{\rm UE}(2,2)$, $\boldsymbol{\mathcal{V}}_{\rm UE}(2,3)$ and $\boldsymbol{\mathcal{V}}_{\rm UE}(2,4)$ are generated using the AMCF-ZCI, enhanced JOINT (EJOINT) and JOINT codeword design schemes, respectively. As a comparison, the ideal beam pattern of $\boldsymbol{\mathcal{V}}_{\rm UE}(2,1)$ is also provided. From Fig.~\ref{BeamPatternCompare}, it is seen that the performance of AMCF-ZCI outperforms that of JOINT and EJOINT. To be specific, EJOINT and JOINT have wider transition band than AMCF-ZCI, since the former two are based on the sub-array combining technique. In particular, the transition band of EJOINT is not monotonous, which may result in the failure of beam training. Moreover, the beam gain of JOINT is lower than AMCF-ZCI because half of antennas are closed for JOINT. Note that AMCF-ZCI can also design codewords with arbitrary beam width, which cannot be achieved by JOINT or EJOINT.

\begin{figure}[!t]
\centering
\includegraphics[width=90mm]{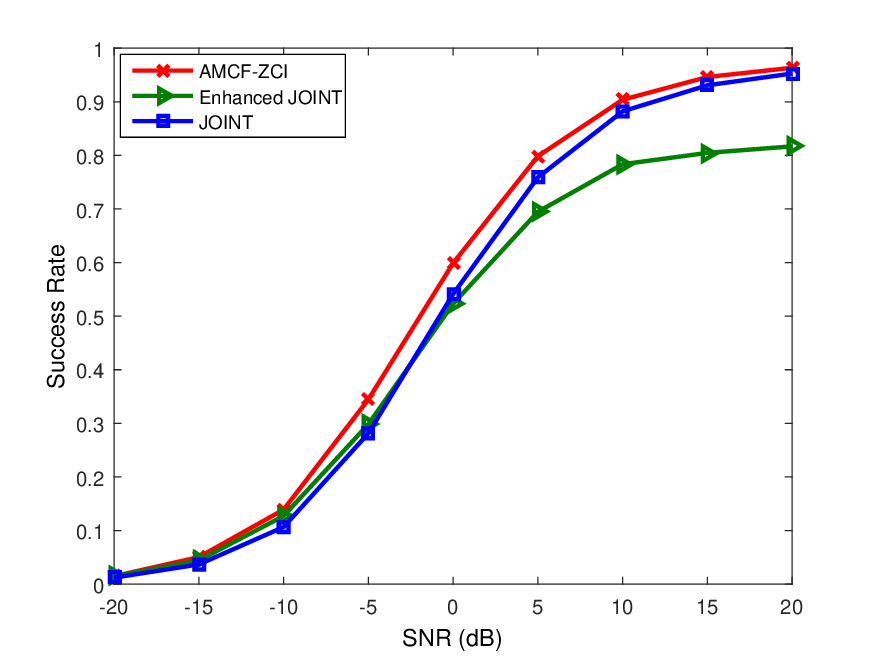}
\caption{Comparisons of beam training performance in terms of success rate using the hierarchical codebooks designed by
different schemes.}
\label{SuccessRate1}
\end{figure}

Now we compare the beam training performance in terms of success rate using the hierarchical codebooks designed by different schemes. The success rate is defined as follows. If the line-of-sight (LOS) path of the UE is correctly identified after beam training, we define that the beam training is successful; otherwise, we define that the beam training is failed. The ratio of the number of successful beam training over the total number of beam training is defined as the success rate. The BS uses the hierarchical codebook designed according to \cite{CKJLETTER}, while the UE uses the codebooks designed by AMCF-ZCI, EJOINT and JOINT, respectively. As shown in Fig.~\ref{SuccessRate1}, we can see that the performance of AMCF-ZCI is better than that of JOINT and EJOINT. The reason on the performance improvement of AMCF-ZCI over JOINT is that AMCF-ZCI uses all the antennas while half of antennas may be powered off by JOINT. The reason on the performance improvement of AMCF-ZCI over EJOINT is that AMCF-ZCI has better beam pattern than EJOINT. Note that EJOINT performs better than JOINT at lower signal-to-noise-ratio (SNR) region and performs worse than JOINT at high SNR region, because EJOINT has a worse beam pattern although it uses all the antennas.

%

\subsection{Evaluation of simultaneous multiuser beam training}
We consider a multiuser mmWave massive MIMO system, where the BS equipped with $N_{\rm BS}=128$ antennas serves $K=8$ UEs. Each UE is equipped with $N_{\rm UE}=16$ antennas. The mmWave MIMO channel matrix is assumed to have $L_k=3$ channel paths with one LOS path and two non-line-of-sight (NLOS) paths, where the channel gain of the LOS path obeys $\lambda_1\sim\mathcal{CN}(0,1)$ and that of the two NLOS paths obeys $\lambda_2\sim\mathcal{CN}(0,0.01)$ and $\lambda_3\sim\mathcal{CN}(0,0.01)$. Both the physical channel AoA $\omega _{\rm UE}^{l}$ and physical channel AoD $\omega _{\rm BS}^{l}$ of the $l$th channel path for $l=1,2,3$ obey the uniform distribution over $[0,2\pi]$~\cite{Sparse2014,TSPMWY}. According to the discussion in the previous subsection, we know that AMCF-ZCI has the best performance among the four schemes taken into comparison. Therefore, we use AMCF-ZCI to design the hierarchical codebook for each UE.


 \begin{figure}[!t]
\centering
\includegraphics[width=90mm]{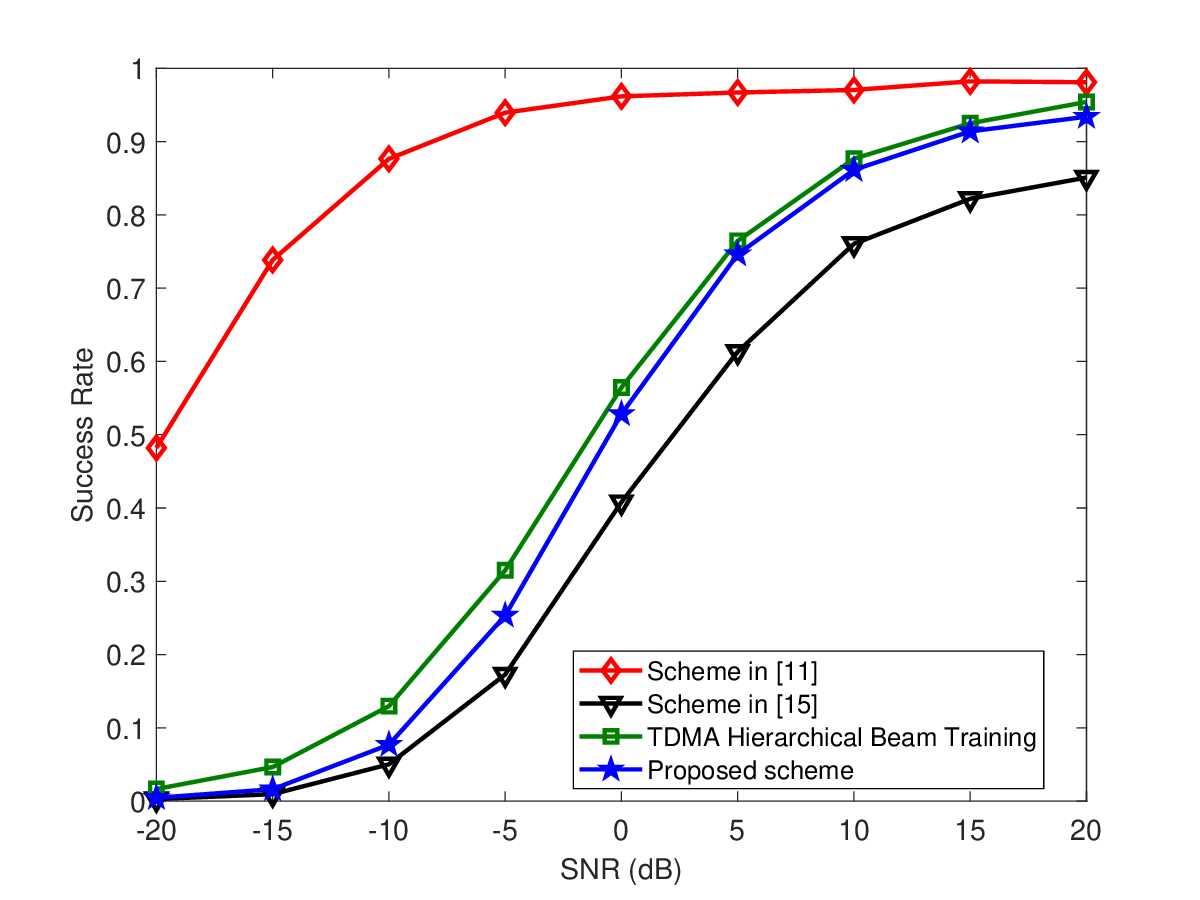}
\caption{Comparisons of success rate of beam training for different schemes.}
\label{Successrate}
\end{figure}

\begin{figure}[!t]
\centering
\includegraphics[width=90mm]{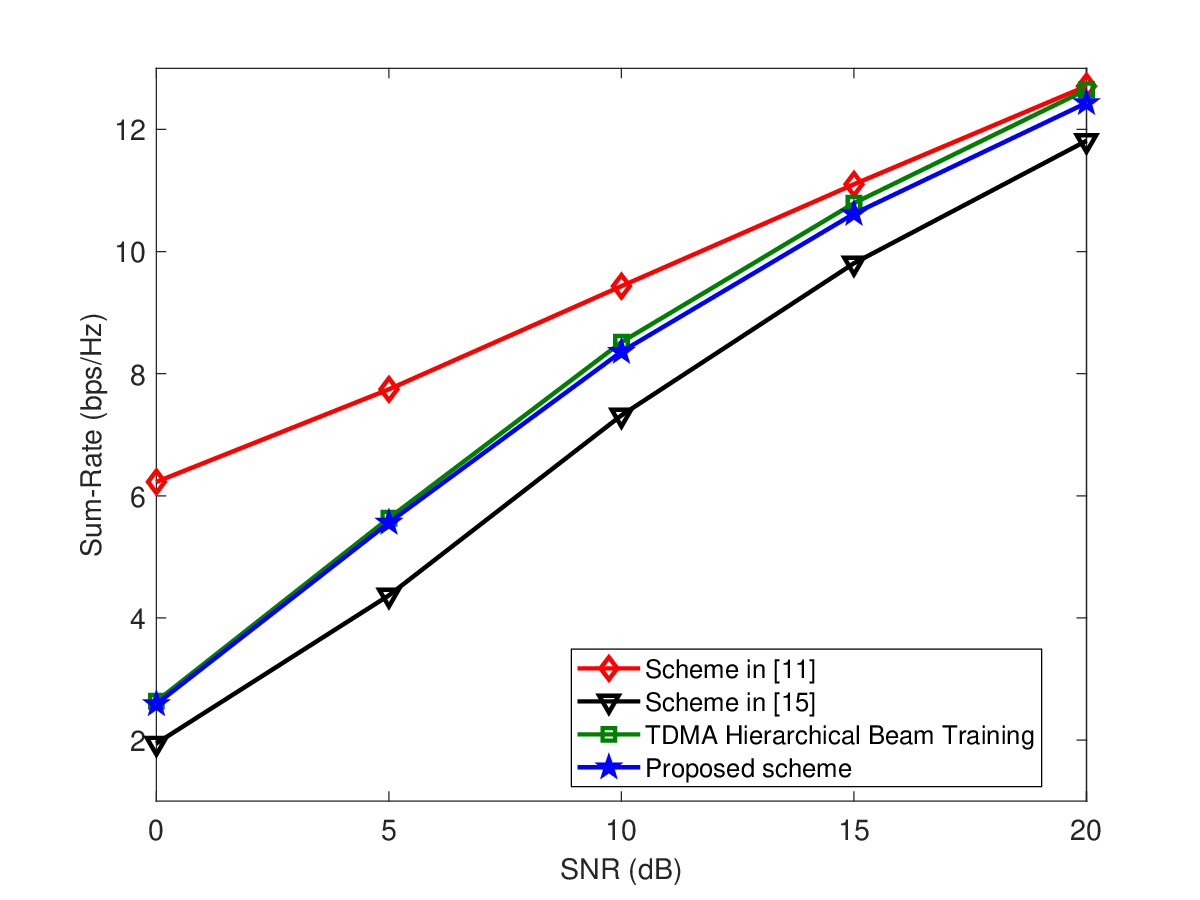}
\caption{Comparisons of the averaged sum-rate for different schemes.}
\label{AchievableRate}
\end{figure}


Fig.~\ref{Successrate} compares the success rate of beam training for different schemes. Since there are totally $K$ UEs served by the BS, the success rate shown in Fig.~\ref{Successrate} is averaged over all $K$ UEs. To make fair comparison, we first extend the scheme in~\cite{SC-2018} from partially connected structure to fully connected structure. It is seen that the scheme in~\cite{AlMultiUser} can achieve better performance than the other three schemes, which lies in the fact that the beam sweeping inherently performs better than the hierarchical beam training. Note that in order to clearly present our idea in this work, we start the hierarchical beam training from the top layer of the hierarchical codebook for both the BS and the UEs. In fact, we may start the hierarchical beam training from the lower layer of the hierarchical codebook to enlarge the beam gain of the codewords, which can improve the beam training performance. Since the training overhead of the scheme in~\cite{AlMultiUser} is much higher than the other schemes, our interest is indeed the comparison of the three hierarchical beam training schemes. It is seen that the performance of our scheme is better than the scheme in~\cite{SC-2018} and almost the same as that of TDMA hierarchical beam training. At low SNR region, e.g., $\textrm{SNR}=-5$dB, our scheme performs slightly worse than the TDMA hierarchical beam training, which is caused by the lower signal power averaged over all UEs in the simultaneous beam training of our scheme. However, the training overhead of our scheme is much smaller than the TDMA hierarchical beam training, i.e., 176 versus 36 with 79.6\% reduction.

Fig.~\ref{AchievableRate} compares the averaged sum-rate for different beam training schemes. It is seen that the curves of our scheme and the TDMA hierarchical beam training scheme are almost overlapped. Moreover, as the SNR increases, the performance gap between our scheme and the scheme in \cite{AlMultiUser} gets smaller. At $\textrm{SNR}=15$dB, the gap is no more than 0.5~bps/Hz, which demonstrates that our scheme can approach the performance of the beam sweeping with the considerable reduction in training overhead.

\section{Conclusion}\label{Sec.Conclusion}
In this paper, we have considered multiuser beam training based on hierarchical codebook for mmWave massive MIMO, where the BS can simultaneously perform the beam training with multiple UEs. For the UEs, we have proposed AMCF-ZCI to design the hierarchical codebook with constant modulus constraint. For the BS, we have designed the hierarchical codebook in an adaptive manner, where the codewords in the current layer are designed according to the beam training results of the previous layer. In particular, we have designed multi-mainlobe codewords for the BS, where each mainlobe of the multi-mainlobe codeword covers a spatial region that one or more UEs are probably in. Except for the bottom layer, there are only two codewords at each layer in the designed adaptive hierarchical codebook, which only requires twice simultaneous beam training for all the UEs no matter how many UEs the BS serves. Simulation results have verified the effectiveness of the proposed hierarchical codebook design schemes and have shown that the proposed simultaneous multiuser beam training scheme can approach the performance of the beam sweeping but with considerable reduction in beam training overhead. Our future work will focus on the reduction of feedback from the UEs to the BS during the multiuser beam training as well as the extension of our beam training scheme from the narrowband mmWave channel model to the wideband one.


\bibliographystyle{IEEEtran}
\bibliography{IEEEabrv,IEEEexample}

@STRING{IEEE_J_COML       = "{IEEE} Commun. Lett."}

@STRING{IEEE_J_JSAC       = "{IEEE} J. Sel. Areas Commun."}

@STRING{IEEE_J_IT         = "{IEEE} Trans. Inf. Theory"}

@ARTICLE{heath2016overview,
	author={Heath, R W and Gonzalez-Prelcic, N and Rangan, S and Roh, W and Sayeed, A},
	journal="{IEEE} J. Sel. Top. Signal Process.",
	title={An Overview of Signal Processing Techniques for Millimeter Wave {MIMO} Systems},
	year={2016},
	volume={10},
	number={3},
	pages={436-453},
	month=Apr,}

@article{Sparse2014,
  author={A. Alkhateeb and O. El Ayach and G. Leus and R. W. Heath},
  journal="{IEEE} J. Sel. Top. Signal Process.",
  title={Channel Estimation and Hybrid Precoding for Millimeter Wave Cellular Systems},
  year={2014},
  volume={8},
  number={5},
  pages={831-846},
  month=Oct,}

@article{PS-DFT2017,
  title={Multi-Resolution Codebook and Adaptive Beamforming Sequence Design for Millimeter Wave Beam Alignment},
  author={S. Noh and M. D. Zoltowski and D. J. Love},
  journal="{IEEE} Trans. Wireless Commun.",
  volume={16},
  number={9},
  pages={5689-5701},
  year={2017},
  month=Sep,}

@article{Xiao2016Hierarchical,
  title={Hierarchical Codebook Design for Beamforming Training in Millimeter-Wave Communication},
  author={Z. Xiao and T. He and P. Xia and X.-G. Xia},
  journal="{IEEE} Trans. Wireless Commun.",
  volume={15},
  number={5},
  pages={3380-3392},
  year={2016},
  month={May.},}

@ARTICLE{CommonCodebookDesign,
	author={J. Song and J. Choi and D. J. Love},
	journal="{IEEE} Trans. Commun.",
	title={Common Codebook Millimeter Wave Beam Design: Designing Beams for Both Sounding and Communication With Uniform Planar Arrays},
	year={2017},
	volume={65},
	number={4},
	pages={1859-1872},
	month=Apr,}

@article{TSPHeShiwen,
  title={Codebook-Based Hybrid Precoding for Millimeter Wave Multiuser Systems},
  author={S. He and J. Wang and Y. Huang and B. Ottersten and W. Hong},
  journal="{IEEE} Trans. Signal Process.",
  volume={65},
  number={20},
  pages={5289-5304},
  year={2017},
  month=Oct,}

@article{TSPMKok,
author={A. Ali and N. Gonzalez-Prelcic and R. W. Heath},
journal="{IEEE} Trans. Wireless Commun.",
title={Millimeter Wave Beam-Selection Using Out-of-Band Spatial Information},
year={2018},
volume={17},
number={2},
pages={1038-1052},
month={Feb.},}

@article{LiYemmWave,
author={C. Lin and G. Y. Li and L. Wang},
journal=IEEE_J_JSAC,
title={Subarray-Based Coordinated Beamforming Training for {mmWave} and Sub-{THz} Communications},
year={2017},
volume={35},
number={9},
pages={2115-2126},
month=Sep,}

@ARTICLE{ZhenyuXiao2018,
author={Z. Xiao and H. Dong and L. Bai and P. Xia and X.-G. Xia},
journal="{IEEE} Trans. Veh. Technol.",
title={Enhanced channel estimation and codebook design for millimeter-wave communication},
year={2018},
volume={67},
number={10},
pages={9393-9405},
month=Oct,
}

@ARTICLE{BLWandYeli,
author={B. Wang and F. Gao and S. Jin and H. Lin and G. Y. Li},
journal="{IEEE} Trans. Signal Process.",
title={Spatial- and Frequency-Wideband Effects in Millimeter-Wave Massive {MIMO} Systems},
year={2018},
volume={66},
number={13},
pages={3393-3406},
month={Jul.},}

@article{TWCSXY,
title={Beam Training and Allocation for Multiuser Millimeter Wave Massive {MIMO} Systems},
author={X. Sun and C. Qi and G. Y. Li},
journal="{IEEE} Trans. Wireless Commun.",
year={2019},
volume={18},
number={2},
pages={1041-1053},
month={Feb.},}

@ARTICLE{TSPMWY,
author={W. Ma and C. Qi},
journal="{IEEE} Trans. Signal Process.",
title={Beamspace Channel Estimation for Millimeter Wave Massive {MIMO} System With Hybrid Precoding and Combining},
year={2018},
volume={66},
number={18},
pages={4839-4853},
month={Sep.},}

@ARTICLE{TSPCKJ,
author={K. Chen and C. Qi and G. Y. Li},
journal="{IEEE} Trans. Signal Process.",
title={Two-Step Codeword Design for Millimeter Wave Massive {MIMO} Systems with Quantized Phase Shifters},
year={2020},
volume={68},
number={1},
pages={170-180},
month={Jan.},}

@inproceedings{SC-2018,
address={Abu Dhabi, UAE},
title={A Codebook Based Simultaneous Beam Training for mmWave Multi-User {MIMO} Systems with Split Structures},
author={R. Zhang and H. Zhang and W. Xu and C. Zhao},
booktitle="2018 {IEEE} Global Commun. Conf. (GLOBECOM)",
pages={1-6},
year={2018},
month=Dec,}

@ARTICLE{CKJLETTER,
author={K. Chen and C. Qi},
journal="{IEEE} Commun. Lett.",
title={Beam Training Based on Dynamic Hierarchical Codebook for Millimeter Wave Massive {MIMO}},
year={2019},
volume={23},
number={1},
pages={132-135},
month={Jan.},}

@ARTICLE{AlMultiUser,
author={A. Alkhateeb and G. Leus and R. W. Heath},
journal="{IEEE} Trans. Wireless Commun.",
title={Limited Feedback Hybrid Precoding for Multi-User Millimeter Wave Systems},
year={2015},
volume={14},
number={11},
pages={6481-6494},
month={Nov.},}

@ARTICLE{ZhaoLou,
author={L. {Zhao} and G. {Geraci} and T. {Yang} and D. W. K. {Ng} and J. {Yuan}},
journal="{IEEE} Trans. Commun.",
title={A Tone-Based {AoA} Estimation and Multiuser Precoding for Millimeter Wave Massive {MIMO}},
year={2017},
volume={65},
number={12},
pages={5209-5225},
month={Dec.},}

@ARTICLE{WZFan,
author={W. Fan and C. Zhang and Y. Huang},
journal="{IEEE} Wireless Commun. Lett.",
title={Flat Beam Design for Massive {MIMO} Systems via Riemannian Optimization},
year={2019},
volume={8},
number={1},
pages={301-304},
month={Feb},
}

@ARTICLE{TractableBeampattern,
author={O. Aldayel and V. Monga and M. Rangaswamy},
journal="{IEEE} Trans. Signal Process.",
title={Tractable Transmit {MIMO} Beampattern Design Under a Constant Modulus Constraint},
year={2017},
volume={65},
number={10},
pages={2588-2599},
month={May}}

@ARTICLE{RuiPeng,
author={R. Peng and Y. Tian},
journal=IEEE_J_COML,
title={Robust Wide-Beam Analog Beamforming With Inaccurate Channel Angular Information},
year={2018},
volume={22},
number={3},
pages={638-641},
month={Mar.},}

@ARTICLE{Zadoff-Chu,
author={D. Chu},
journal=IEEE_J_IT,
title={Polyphase codes with good periodic correlation properties (Corresp.)},
year={1972},
volume={18},
number={4},
pages={531-532},
month={Jul.},}

@article{xiao2018noma,
  title={Joint Power Allocation and Beamforming for Non-Orthogonal Multiple Access {(NOMA)} in {5G} Millimeter-Wave Communications},
  author={Xiao, Z and Zhu, Lipeng and Choi, Jinho and Cao, Xianbin and Xia, Xiang-Gen},
  journal={IEEE Trans. Wireless Commun.},
  volume={17},
  number={5},
  pages={2961--2974},
  year={2018},
  month=May,
}

@article{Ma2020Sparse,
  title={Sparse Channel Estimation and Hybrid Precoding Using Deep Learning for Millimeter Wave Massive {MIMO}},
  author={Ma, Wenyan and Qi, Chenhao and Zhang, Zaichen and Cheng, Julian},
  journal={IEEE Trans. Commun.},
  volume={68},
  number={5},
  pages={2838-2849},
  month={May.},
  year={2020},
}

@article{Wei2019multi,
  title={Multi-Beam {NOMA} for Hybrid {mmWave} Systems},
  author={Wei, Zhiqiang and Zhao, Lou and Guo, Jiajia and Derrick Wing Kwan Ng and Yuan, Jinhong},
  journal={IEEE Trans. Commun.},
  volume={67},
  number={2},
  pages={1705-1719},
  month={Feb.},
  year={2019},
}

@article{ChenhaoQiOverview,
  title={Acquisition of Channel State Information for {mmWave} Massive {MIMO}: {Traditional} and Machine Learning-based Approaches},
  author={Qi, Chenhao and Dong, Peihao and Ma, Wenyan and Zhang, Hua and Zhang Zaichen and Li, Geoffrey Ye},
  journal={arXiv:2006.08894},
  month={June},
  year={2020},
}

@inproceedings{chen2019simultaneous,
  address={Waikoloa Village, HI, USA},
  title={Simultaneous Multiuser Beam Training Using Adaptive Hierarchical Codebook for mmWave Massive {MIMO}},
  author={Chen, Kangjian and Qi, Chenhao and Dobre, Octavia A and Li, Geoffrey Ye},
  booktitle={2019 {IEEE} Global Commun. Conf. (GLOBECOM)},
  pages={1-6},
  year={2019},
  month=Dec,
}

\end{document}